\documentclass{article}

\usepackage{arxiv}

\usepackage[utf8]{inputenc} 
\usepackage[T1]{fontenc}    
\usepackage{hyperref}       
\usepackage{url}            
\usepackage{booktabs}       
\usepackage{amsfonts}       
\usepackage{nicefrac}       
\usepackage{microtype}      
\usepackage{lipsum}
\usepackage{graphicx}

\title{Determining feature importance for actionable climate change mitigation policies}

\author{
  Romit Maulik\thanks{Corresponding author} \\
  Argonne Leadership Computing Facility\\
  Argonne National Laboratory\\
  Lemont, IL-60439 \\
  \texttt{rmaulik@anl.gov} \\
  \And
  Junghwa Choi \\
  Department of Political Science\\
  University of Oklahoma\\
  Norman, OK-73019 \\
  \texttt{junghwa.choi-1@ou.edu} \\
  \And
  Wesley Wehde \\
  Department of Political Science, International Affairs, and Public Administration\\
  East Tennessee State University\\
  Johnson City, TN-37614 \\
  \texttt{wehdew@etsu.edu} \\
  \And
  Prasanna Balaprakash \\
  Mathematics and Computer Science \& Argonne Leadership Computing Facility \\
  Argonne National Laboratory\\
  Lemont, IL-60439 \\
  \texttt{pbalapra@anl.gov} \\
}

\begin{document}
\maketitle

\begin{abstract}
Given the importance of public support for policy change and implementation, public policymakers and researchers have attempted to understand the factors associated with this support for climate change mitigation policy. In this article, we compare the feasibility of using different supervised learning methods for regression using a novel socio-economic data set which measures public support for potential climate change mitigation policies. Following this model selection, we utilize gradient boosting regression, a well-known technique in the machine learning community, but relatively uncommon in public policy and public opinion research, and seek to understand what factors among the several examined in previous studies are most central to shaping public support for mitigation policies in climate change studies. The use of this method provides novel insights into the most important factors for public support for climate change mitigation policies. Using national survey data, we find that the perceived risks associated with climate change are more decisive for shaping public support for policy options promoting renewable energy and regulating pollutants. However, we observe a very different behavior related to public support for increasing the use of nuclear energy where climate change risk perception is no longer the sole decisive feature. Our findings indicate that public support for renewable energy is inherently different from that for nuclear energy reliance with the risk perception of climate change, dominant for the former, playing a subdued role for the latter.
\end{abstract}


\section{Introduction}
Scientific evidence has shown that global climate change is happening at an unprecedented rate in recent decades \cite{stocker2013climate}. Climate scientists have reported that climate change is heavily human-induced, and it poses a serious threat to humans and natural ecosystems  \cite{stocker2013climate}. Even worse, it is expected that the damages and related costs of climate change are likely to increase over time \cite{stocker2013climate} with some forecasts estimating over \$200 billion per year in costs spread over 22 different economic sectors \cite{martinich2019climate}. Given these circumstances, scholars and policymakers have sought to develop a wide range of mitigation policy options that address the causes of and risks associated with climate change. 

However, public support for climate change mitigation policies remains equivocal while the scientific evidence for climate change is not \cite{lee2015predictors}. Scholars of political science and public attitudes towards climate change have argued that public support for public policy is a key determinant of policy adoption in democratic societies \cite{lee2015predictors,ziegler2017political}. Furthermore, successful implementation of public policy can be facilitated when the public supports the policy options \cite{sabatier1980implementation}. Given the urgency of addressing climate change, lack of public support for climate change mitigation policies may become a major blockade for adopting and implementing climate change mitigation policies. Therefore, understanding public support for climate mitigation policies and its associated determinants can be tremendously valuable for decision-makers when deciding and implementing specific policy options to mitigate the risks associated with climate change. 

Given the importance of understanding public support for climate change mitigation policy, drawing theories from various disciplines, scholars have sought to understand the determinants and motivations of individual support for climate change mitigation policy \cite{drews2016explains}. Scholars have found that demographic characteristics, political dispositions and issue-specific factors such as social-psychological factors and climate change risk perceptions are associated with public support for climate change mitigation policy through traditional regression approaches \cite{tobler2012addressing, o1999risk, lubell2007collective}. However, to the best of our knowledge, public policy scholars have not attempted to answer the question of relative decisiveness of factors shaping public support for climate change mitigation policies and have remained limited to linear parametric methods alone. Therefore, we ask, among the factors previously investigated, what factors are the most decisive in structuring public support for climate change mitigation policies? 

To answer this question, we examine the prediction ability of multiple nonlinear data-driven techniques to ascertain the optimal modeling strategy for our data set. Following this model selection step, we settle on the use of gradient boosting regressors to answer our question. This non-parametric and data-driven technique allows us to calculate the relative decisiveness of explanatory variables among many factors. Besides, it relaxes the linear and uncorrelated assumption between features and targets associated with linear and ridge regression-based models to allow for the modeling of rich nonlinear relationships.

In the rest of the paper, first, we review previous literature of public support for the selection of explanatory variables. We then describe our data, from the 2019 Oklahoma Meso-Scale Integrated Socio-geographic Network National Comparison Survey. This survey includes a set of questions regarding individual support for climate change mitigation policies in addition to several climate change issue-specific questions, political dispositions as well as basic demographic characteristics. As dependent variables – given by individual support for climate change mitigation policy, we specifically look at individual support for 1) general intervention of federal government to mitigate risks associated with climate change 2) three policy options highlighting renewable energy and 3) nuclear energy as a way to reduce greenhouse gas emissions. Our primary findings are as follows:
\begin{itemize}
    \item We find that the risk perception of climate change and belief in the link between climate change and greenhouse gas emissions are the most decisive factors for structuring individual support for government intervention and policy options related to renewable energy.
    \item We find a different relationship concerning individual support for the nuclear energy option with no stand-out features that are more important than the others, thereby indicating that this policy initiative cannot be categorized with other renewable energy-based mitigation strategies.
\end{itemize}


\section{Literature Review}
Scholars in various disciplines including political science, public policy, psychology, among others have utilized linear regression methods and sought to understand individual determinants of support for environmental and climate change policies. Individual policy support and preferences are complex, therefore, there are many potential explanatory determinants. Recently, based on a wide range of previous studies on individual policy support and preference, scholars have found that three sets of determinants, namely, basic demographic characteristics, political (and social) predispositions, and issue-specific factors are closely associated with individual policy support and preference \cite{robinson2017understanding, choi2019venue}. 

The first set of determinants is individual demographic characteristics. They typically considered to account for potential confounding effects on individual policy support and preference in statistical modeling. However, scholars have found that individual attributes are often more stable determinants and predictors of individual support and preferences for environmental policy than other social determinants \cite{jones1992social}. While the directions, strengths, and significance of relationships are often contingent upon the specific policy area under consideration \cite{liurobinsonvedlitz2017}, scholars have found that gender, income, location of residence, age, number of children at household, employment and education are generally associated with individual policy support and preferences \cite{boudet2014fracking}. 

The second set of determinants is political and social predispositions. Political and social predispositions are particular attitudes or beliefs towards political and social phenomena through which individuals process policy options \cite{rudolph2005political}. This set of variables typically includes political ideology, party identification, public trust, religiosity and worldview and has been highlighted as determinants to shape individual policy support and preference across a wide variety of domains including climate change mitigation policy \cite{reckhow2015, mumpower2016predictors}. For instance, scholars have found that party identification and political ideology, particularly Democratic party affiliation and liberal political orientation are strongly related to policy support among individuals \cite{mccright2011politicization,mccright2013perceived}. Furthermore, previous studies have indicated that worldviews, operationalized through cultural theory, are a strong determinant of individual policy support and preferences \cite{Stoutenborough_Sturgess_Vedlitz_2013}. More specifically, those who hold egalitarian worldviews are more likely to support climate change mitigation policy while individualistic and hierarchical individuals do not tend to support climate change mitigation policy \cite{smith2013american}. Scholars have also found that public trust matters. It has been observed that social trust, defined as general trust in others in their community and trust in government(s) are strongly associated with individual attitudes and behaviors towards risk mitigation policy including climate change mitigation policy (see \cite{drews2016explains, choitrust} and references therein). Finally, scholars have also argued that religiosity and general beliefs about the human-nature relationship are particularly important to understanding individual support and preference for climate change mitigation policy \cite{smith2013american}. 
The last set of determinants is issue-specific factors. These variables generally include individual attention, perceptions about climate change and personal experience. Issue attention is defined as the extent to which individuals show interest in a policy issue or problem. Scholars have measured issue attention by looking at individual Google search patterns, concerns about energy supply in relation to environmental issues, and the use of news media \cite{liurobinsonvedlitz2017, holt2013age}. More specifically, in environmental policy, scholars have found that general environmental concerns are associated with more support for specific environmental protection and climate change mitigation policies \cite{liurobinsonvedlitz2017}. Previous studies have also shown that individuals who directly or indirectly related risks have higher risk perceptions of climate change and are highly likely to support policies mitigating those issues \cite{spence2011perceptions}. Beliefs about the causality of issues, such as climate change, are also strongly related to individual policy \cite{ding2011support, choi2019venue}. Risk perception has been heavily highlighted as a determinant of individual support for mitigation policies \cite{Leiserowitz_2006,maestas2018fearing}. For instance, \cite{choi2019venue} find that when individuals perceive higher risks associated with earthquakes, they are more likely to support government mitigation policy. Finally, previous literature has demonstrated that issue causality shapes individual policy support and preference. Scholars have found that people who define climate change as a human-induced hazard tend to support climate change mitigation policy more than those who do not \cite{leiserowitz2006climate, brouwer2008convenient, krosnick2006origins}.

\section{Data}
\label{Data}

The survey data for this project is open-source (Available at \texttt{http\-://crcm.ou.edu/epscordata/}) and were collected using a national internet panel sample from Qualtrics with quotas for gender, race, ethnicity (Hispanic), age, and Census region. Data were collected between September 5th and September 30th, 2018. The survey included a variety of questions related to weather, climate, and energy policy with a final sample size of 1,800 respondents. All responses are used for the purpose of our analyses. Our dependent variables measure support for a variety of policies that seek to address climate change. Specific question wordings can be found \href{http://crcm.ou.edu/epscor/codebooks/codebook-wave19-national.pdf}{here}; responses for specific policy questions are on a scale from strongly oppose (1) and somewhat oppose (2) to somewhat support (3) and strongly support (4). The four specific policies we examine are increasing funding for renewable energy research, generating renewable energy on public land, regulating CO2 as a pollutant, and increasing nuclear energy reliance. General policy involvement in climate change mitigation is measured on a scale from 0, representing not at all involved, to 10, representing extremely involved, for the federal government. This leads to a total of five dependent variables for which we attempt to characterize feature importance.

We include a large selection of features drawn from demographic characteristics, political and social identities and belief systems, and climate change and environmental specific beliefs. For demographics, we include measures for gender (1 = male), age (in years), ethnicity (1 = Hispanic), race (1 = white), education (ranging from 1 for less than high school to 8 for doctorate education), employment (1 = working), income (in dollars), household location (1 = rural) and number of children in the household. For political and social dispositions, we include measures of partisanship (1 = Democrat), ideology (1 = Strongly liberal to  7 = Strongly conservative), cultural worldview (1 = hierarchs, 2 = individualists, 3 = egalitarians, and 4 = fatalists), social trust or capital (based on the average of six measures scaled from 1 for strongly disagree to 7 for strongly agree), importance of religion (from 0 = not at all to 10 = extremely important), and human dominion over nature (average of two items ranging from 1 = strongly disagree to 5 = strongly agree). Finally, for environmental and climate change specific attributes, we include measures of perceptions of future extreme heat events (1 = less often, 2 = about the same, and 3 = more often), an index of energy-saving actions (from 0 = none taken to 8 = all taken), general concern for natural resource preservation (from 0 = not at all to 10 = extremely concerned), cause of climate change (1 = caused by greenhouse gas emissions), and climate change risk perceptions (from 0 = no risk to 10 = extreme risk). 

\section{Results}

Disciplines such as political science, public policy, psychology, among others have frequently utilized linear regression methods and sought to understand individual determinants of support for environmental and climate change policies. The existing literature focuses on regression-based models that seek to identify the difference in expected levels of preparedness (the dependent variable) based on ensemble lists of suspected factors related to preparedness (all of the features). This approach provides information about the statistical significance of proposed factors and, sometimes, an effect size related to a variable under specific parametric assumptions. The parametric assumptions become quite strong as one includes interrelated features -- as is almost universally the case. We seek an alternative model, instead, to assess the decisiveness of the variable rather than an expected difference.  This search led us to a class of models relatively new to studies of climate change attitudes and beliefs - ensemble machine learning models. Our workflow relies on an exhaustive assessment of nonlinear modeling strategies following by using the best possible model (based on predictive ability) to obtain feature decisiveness. Following an assessment of predictive ability, the best model is utilized for providing feature importance based on explained variance. The various methods assessed here are LR: linear regression, RR: Ridge regression, SVM: Support vector machine, GP: Gaussian Process, KNN: K-nearest neighbors, DT: Decision tree, BR: Bagging regression, ETR: Extra-trees regression, RFR: Random-forest regression, ABR: AdaBoost Regression, GBR: Gradient boost regression. We note that the methods utilized here are based on Scikit-learn \cite{scikit_learn} defaults and GBR uses the well-known package XGBoost \cite{ChenXGB} (further details can be observed in supporting code). Readers may find source-code and data for the reproduction of results at \texttt{https://github.com/Romit-Maulik/Climate\_Change\_Importance}.

We undertake an analysis of the data set introduced in Section \ref{Data} by building predictive models for each of the dependent variables using our selection of linear and nonlinear data-driven methods. Following this, we identify the best predictive model through the use of metrics before moving forward with variable importance assessments. We utilize four standard metrics on the test data set given by $R^2$ (the coefficient of determination), $\rho$ (the correlation coefficient), EVS (the explained variance in the data), MAE (the mean absolute error) and RMSE (the root-mean-squared error) to settle on a choice for the final model all computed from the Scikit-learn metrics module. These assessments are performed on all the dependent variables corresponding to responses for specific policy questions as outlined in Section \ref{Data} and use 40-fold cross-validation where each fold was generated by a random 80\% training and 20\% testing split of the total data. In the following, we designate the metrics for one fit in the sequence [$R^2$, $\rho$, EVS, MAE, RMSE].

Table \ref{Table_1} shows results from an assessment on the dependent variable corresponding to support for federal involvement in climate change mitigation with GBR providing the best results (a mean of [0.487, 0.699, 0.488, 1.528, 2.181] respectively). We also draw attention to the performance of linear methods which show very high errors on this dataset (mean metrics for LR - [-2.12E+23, 0.56, -2.12E+23, 3.15E+10, 6.36E+11] and RR - [-2.12E+23, 0.56, -2.12E+23, 3.15E+10, 6.36E+11]). Among the other methods tested, SVMs show the second-best performance comparable to the GBR (with mean metrics of [0.47, 0.698, 0.479, 1.514, 2.215]). Both GBR and SVM are significantly better than the other alternatives with the next nearest method (in terms of accuracy) being the RFR (mean metrics of [0.422,0.658, 0.425, 1.641, 2.314]). Each table also shows the standard deviation of all these metrics to identify methods exhibiting large deviations in accuracy. The GBR and the SVM, once again, show comparable metrics in terms of standard deviation (GBR - [0.0486, 0.0332, 0.0482, 0.0661, 0.121], SVM - [0.0505, 0.0314, 0.0488, 0.0748, 0.122]).

Similar results for the other policy options are outlined in tables \ref{Table_3} (support for renewable research funding), \ref{Table_4} (support for generating renewable energy on public land), \ref{Table_5} (support for CO2 emission regulation). These tables also outline that the GBR is seen to perform the most accurately with $R^2$ values of 0.193, 0.168 and 0.271 respectively. SVM is seen to be the next most accurate model with $R^2$ values of 0.167, 0.102 and 0.248 respectively. Note, however, that the SVM occasionally obtains better mean MAE values than the GBR (for instance in Tables \ref{Table_1}, \ref{Table_3}, \ref{Table_4} and \ref{Table_5}). Also observed is the fact that LR and RR perform very poorly for all the policy options reflecting the limitations of a linear uncorrelated assumption for the features.

We would like to caution the reader here that most of these (particularly $R^2$ and EVS) are significantly lower than for mainstream machine learning applications due to the psychological nature of the study. This is the primary reason we evaluate $R^2$ values in conjunction with other residual quantities and subject area knowledge in order to obtain more confident conclusions. Table \ref{Table_2} outlines metrics obtained by fitting the various models for the policy option related to support for nuclear energy reliance where very low $R^2$ values (almost all less than or slightly above zero) are obtained (see Table \ref{Table_2}). The GBR is the only method that can obtain a mean positive $R^2$ value given by 0.000347. Note, however, that the standard deviation for this value is  (a much greater) 0.028. The poor performance of all models (linear or otherwise) across the board hints at the features not being able to capture any meaningful relationship with the target policy option.

\begin{table*}[h]
\centering
\small
\caption{An assessment of predictive ability for one of our dependent variables (support for federal involvement) with mean metrics (top) and corresponding standard deviations (bottom). Our nomenclature is given by: LR: linear regression, RR: Ridge regression, SVM: Support vector machine, GP: Gaussian Process, KNN: K-nearest neighbors, DT: Decision tree, BR: Bagging regression, ETR: Extra-trees regression, RFR: Random-forest regression, ABR: AdaBoost Regression, GBR: Gradient boosting regression. GBR is seen to provide most accurate results.}
\begin{tabular}{|c|c|c|c|c|c|c|c|c|c|c|c|}
\hline
\multicolumn{1}{|c|}{} & \multicolumn{11}{c|}{Methods (mean)}                                                                                                  \\ \hline
               & LR          & RR          & SVM    & GP     & KNN    & DT       & BR     & ETR    & RFR    & ABR    & GBR             \\ \hline
$R^2$                  & -2.12E+023 & -2.12E+023 & 0.4707 & 0.1546 & 0.3772 & -0.05094 & 0.4146 & 0.4025 & 0.4227 & 0.333  & \textbf{0.487}  \\ \hline
$\rho$                 & 0.5555      & 0.5555      & 0.6986 & 0.4544 & 0.6221 & 0.4863   & 0.6527 & 0.648  & 0.6579 & 0.6651 & \textbf{0.6995} \\ \hline
EVS                    & -2.12E+023 & -2.12E+023 & 0.4796 & 0.1581 & 0.3793 & -0.0475  & 0.4171 & 0.405  & 0.4248 & 0.4048 & \textbf{0.4882} \\ \hline
MAE                    & 3.15E+10    & 3.15E+10    & 1.514  & 2.243  & 1.772  & 2.076    & 1.658  & 1.67   & 1.641  & 2.032  & \textbf{1.528}  \\ \hline
RMSE                   & 6.36E+11    & 6.36E+11    & 2.215  & 2.804  & 2.403  & 3.122    & 2.33   & 2.355  & 2.314  & 2.486  & \textbf{2.181}  \\ \hline
\end{tabular}

\vspace{0.4cm}

\begin{tabular}{|c|c|c|c|c|c|c|c|c|c|c|c|}
\hline
            & \multicolumn{11}{c|}{Methods (St. Dev.)}                                                                                            \\ \hline
 & LR        & RR        & SVM    & GP     & KNN    & DT     & BR     & ETR    & RFR    & ABR    & GBR             \\ \hline
$R^2$       & 4.28E+023 & 4.28E+023 & 0.0505 & 0.0162 & 0.0554 & 0.0883 & 0.0478 & 0.0432 & 0.049  & 0.0668 & \textbf{0.0486} \\ \hline
$\rho$      & 0.289     & 0.289     & 0.0314 & 0.0293 & 0.0389 & 0.0392 & 0.0314 & 0.0278 & 0.0326 & 0.0342 & \textbf{0.0332} \\ \hline
EVS         & 4.27E+023 & 4.27E+023 & 0.0488 & 0.0152 & 0.055  & 0.0881 & 0.0471 & 0.0426 & 0.0485 & 0.0386 & \textbf{0.0482} \\ \hline
MAE         & 6.32E+10  & 6.32E+10     & 0.0748 & 0.0848 & 0.0717 & 0.108  & 0.0734 & 0.0674 & 0.0727 & 0.141  & \textbf{0.0661} \\ \hline
RMSE        & 1.27E+12  & 1.27E+12     & 0.122  & 0.102  & 0.114  & 0.136  & 0.11   & 0.0932 & 0.109  & 0.121  & \textbf{0.121}  \\ \hline
\end{tabular}
\label{Table_1}
\end{table*}

\begin{table*}[h!]
\centering
\small
\caption{An assessment of predictive ability for one of our dependent variables (support for increasing funding for renewable research) with mean metrics (top) and corresponding standard deviations (bottom). Our nomenclature is given by: LR: linear regression, RR: Ridge regression, SVM: Support vector machine, GP: Gaussian Process, KNN: K-nearest neighbors, DT: Decision tree, BR: Bagging regression, ETR: Extra-trees regression, RFR: Random-forest regression, ABR: AdaBoost Regression, GBR: Gradient boosting regression. GBR is seen to provide most accurate results.}
\begin{tabular}{|c|c|c|c|c|c|c|c|c|c|c|c|}
\hline
       & \multicolumn{11}{c|}{Methods (Mean) }                                                                                 \\ \hline
  & LR         & RR         & SVM   & GP     & KNN    & DT     & BR     & ETR    & RFR   & ABR   & GBR            \\ \hline
$R^2$  & -4.46E+023 & -4.45E+023 & 0.167 & 0.0545 & 0.06   & -0.612 & 0.0996 & 0.0838 & 0.102 & 0.117 & \textbf{0.193} \\ \hline
$\rho$ & 0.359      & 0.359      & 0.438 & 0.248  & 0.339  & 0.233  & 0.384  & 0.374  & 0.386 & 0.414 & \textbf{0.449} \\ \hline
EVS    & -4.45E+023 & -4.44E+023 & 0.182 & 0.0582 & 0.0629 & -0.608 & 0.103  & 0.0876 & 0.106 & 0.168 & \textbf{0.196} \\ \hline
MAE    & 1.12E+10   & 1.12E+10   & 0.528 & 0.643  & 0.585  & 0.677  & 0.571  & 0.575  & 0.57  & 0.613 & \textbf{0.539} \\ \hline
RMSE   & 2.45E+11   & 2.45E+11   & 0.741 & 0.79   & 0.787  & 1.03   & 0.77   & 0.777  & 0.769 & 0.762 & \textbf{0.729} \\ \hline
\end{tabular}

\vspace{0.4cm}

\begin{tabular}{|c|c|c|c|c|c|c|c|c|c|c|c|}
\hline
       & \multicolumn{11}{c|}{Methods (St. Dev.)}                                                                                   \\ \hline
   & LR        & RR        & SVM    & GP     & KNN    & DT     & BR     & ETR    & RFR    & ABR    & GBR             \\ \hline
$R^2$  & 9.01E+023 & 9E+023    & 0.034  & 0.0155 & 0.0474 & 0.133  & 0.0518 & 0.0557 & 0.0512 & 0.0469 & \textbf{0.0394} \\ \hline
$\rho$ & 0.222     & 0.222     & 0.0334 & 0.0401 & 0.0396 & 0.0382 & 0.0413 & 0.0407 & 0.037  & 0.0333 & \textbf{0.0363} \\ \hline
EVS    & 8.99E+023 & 8.97E+023 & 0.0356 & 0.0155 & 0.0469 & 0.132  & 0.0493 & 0.0533 & 0.0484 & 0.0246 & \textbf{0.0383} \\ \hline
MAE    & 2.43E+10  & 2.43E+10  & 0.027  & 0.025  & 0.0252 & 0.0379 & 0.027  & 0.0223 & 0.0215 & 0.0259 & \textbf{0.0226} \\ \hline
RMSE   & 4.90E+11  & 4.90E+11  & 0.0392 & 0.0362 & 0.0331 & 0.0384 & 0.034  & 0.0341 & 0.03   & 0.0279 & \textbf{0.033}  \\ \hline
\end{tabular}
\label{Table_3}
\end{table*}

\begin{table*}[ht]
\centering
\small
\caption{An assessment of predictive ability for one of our dependent variables (support for generating renewable energy on public land) with mean metrics (top) and corresponding standard deviations (bottom). Our nomenclature is given by: LR: linear regression, RR: Ridge regression, SVM: Support vector machine, GP: Gaussian Process, KNN: K-nearest neighbors, DT: Decision tree, BR: Bagging regression, ETR: Extra-trees regression, RFR: Random-forest regression, ABR: AdaBoost Regression, GBR: Gradient boosting regression. GBR is seen to provide most accurate results.}
\begin{tabular}{|c|c|c|c|c|c|c|c|c|c|c|c|}
\hline
       & \multicolumn{11}{c|}{Methods (Mean) }                                                                                    \\ \hline
  & LR         & RR         & SVM   & GP     & KNN     & DT     & BR     & ETR    & RFR    & ABR    & GBR            \\ \hline
$R^2$  & -4.01E+023 & -4.01E+023 & 0.102 & 0.0405 & 0.00259 & -0.72  & 0.0555 & 0.0166 & 0.0566 & 0.0224 & \textbf{0.168} \\ \hline
$\rho$ & 0.307      & 0.307      & 0.378 & 0.212  & 0.283   & 0.189  & 0.337  & 0.311  & 0.34   & 0.345  & \textbf{0.42}  \\ \hline
EVS    & -4E+023    & -4E+023    & 0.125 & 0.0437 & 0.00558 & -0.714 & 0.0599 & 0.0211 & 0.0608 & 0.118  & \textbf{0.17}  \\ \hline
MAE    & 1.15E+10   & 1.15E+10   & 0.547 & 0.649  & 0.606   & 0.704  & 0.591  & 0.599  & 0.591  & 0.652  & \textbf{0.558} \\ \hline
RMSE   & 2.32E+11   & 2.32E+11   & 0.767 & 0.793  & 0.808   & 1.06   & 0.786  & 0.802  & 0.786  & 0.799  & \textbf{0.738} \\ \hline
\end{tabular}

\vspace{0.4cm}

\begin{tabular}{|c|c|c|c|c|c|c|c|c|c|c|c|}
\hline
       & \multicolumn{11}{c|}{Methods (St. Dev.)}                                                                                   \\ \hline
   & LR        & RR        & SVM    & GP     & KNN    & DT     & BR     & ETR    & RFR    & ABR    & GBR             \\ \hline
$R^2$  & 8.12E+023 & 8.1E+023  & 0.0439 & 0.0133 & 0.0536 & 0.18   & 0.0573 & 0.0589 & 0.0507 & 0.084  & \textbf{0.0485} \\ \hline
$\rho$ & 0.227     & 0.227     & 0.0437 & 0.034  & 0.043  & 0.0507 & 0.0451 & 0.0431 & 0.0383 & 0.0474 & \textbf{0.0481} \\ \hline
EVS    & 8.1E+023  & 8.08E+023 & 0.0457 & 0.0124 & 0.0533 & 0.176  & 0.0545 & 0.0568 & 0.0498 & 0.0327 & \textbf{0.0478} \\ \hline
MAE    & 2.31E+10  & 2.31E+10  & 0.0225 & 0.0198 & 0.0215 & 0.0329 & 0.0228 & 0.0194 & 0.0212 & 0.0281 & \textbf{0.0205} \\ \hline
RMSE   & 4.65E+11  & 4.65E+11  & 0.0343 & 0.0304 & 0.0274 & 0.0419 & 0.0297 & 0.0296 & 0.0286 & 0.0299 & \textbf{0.0298} \\ \hline
\end{tabular}
\label{Table_4}
\end{table*}

\begin{table*}[ht]
\centering
\small
\caption{An assessment of predictive ability for one of our dependent variables (support for regulating CO2 as a pollutant) with mean metrics (top) and corresponding standard deviations (bottom). Our nomenclature is given by: LR: linear regression, RR: Ridge regression, SVM: Support vector machine, GP: Gaussian Process, KNN: K-nearest neighbors, DT: Decision tree, BR: Bagging regression, ETR: Extra-trees regression, RFR: Random-forest regression, ABR: AdaBoost Regression, GBR: Gradient boosting regression. GBR is seen to provide most accurate results.}
\label{Table_5}
\begin{tabular}{|c|c|c|c|c|c|c|c|c|c|c|c|}
\hline
       & \multicolumn{11}{c|}{Methods (Mean)}                                                                              \\ \hline
   & LR         & RR         & SVM   & GP     & KNN   & DT     & BR    & ETR   & RFR   & ABR   & GBR            \\ \hline
$R^2$  & -4.82E+023 & -4.81E+023 & 0.248 & 0.0929 & 0.171 & -0.514 & 0.189 & 0.169 & 0.198 & 0.205 & \textbf{0.271} \\ \hline
$\rho$ & 0.422      & 0.422      & 0.52  & 0.335  & 0.452 & 0.284  & 0.471 & 0.46  & 0.476 & 0.49  & \textbf{0.527} \\ \hline
EVS    & -4.81E+023 & -4.8E+023  & 0.256 & 0.097  & 0.174 & -0.508 & 0.192 & 0.172 & 0.201 & 0.222 & \textbf{0.273} \\ \hline
MAE    & 1.42E+10   & 1.42E+10   & 0.594 & 0.674  & 0.64  & 0.771  & 0.623 & 0.623 & 0.62  & 0.645 & \textbf{0.592} \\ \hline
RMSE   & 2.87E+11   & 2.87E+11   & 0.795 & 0.873  & 0.834 & 1.13   & 0.825 & 0.835 & 0.82  & 0.817 & \textbf{0.782} \\ \hline
\end{tabular}

\vspace{0.4cm}

\begin{tabular}{|c|c|c|c|c|c|c|c|c|c|c|c|}
\hline
       & \multicolumn{11}{c|}{Methods (St. Dev.)}                                                                                   \\ \hline
   & LR        & RR        & SVM    & GP     & KNN    & DT     & BR     & ETR    & RFR    & ABR    & GBR             \\ \hline
$R^2$  & 9.7E+023  & 9.68E+023 & 0.0461 & 0.0183 & 0.0482 & 0.148  & 0.0515 & 0.0489 & 0.0559 & 0.0337 & \textbf{0.0479} \\ \hline
$\rho$ & 0.243     & 0.243     & 0.0366 & 0.0409 & 0.0376 & 0.0552 & 0.038  & 0.0356 & 0.0418 & 0.0373 & \textbf{0.0398} \\ \hline
EVS    & 9.68E+023 & 9.66E+023 & 0.0467 & 0.0173 & 0.0466 & 0.145  & 0.0501 & 0.0478 & 0.055  & 0.0268 & \textbf{0.0474} \\ \hline
MAE    & 2.85E+10  & 2.85E+10  & 0.0211 & 0.0224 & 0.0224 & 0.0425 & 0.021  & 0.024  & 0.0209 & 0.0207 & \textbf{0.0211} \\ \hline
RMSE   & 5.75E+11  & 5.75E+11  & 0.0294 & 0.0246 & 0.0269 & 0.05   & 0.0286 & 0.0286 & 0.0288 & 0.022  & \textbf{0.0282} \\ \hline
\end{tabular}
\end{table*}

\begin{table*}[h]
\centering
\small
\caption{An assessment of predictive ability for one of our dependent variables (support for increased nuclear energy reliance) with mean metrics (top) and corresponding standard deviations (bottom). Our nomenclature is given by: LR: linear regression, RR: Ridge regression, SVM: Support vector machine, GP: Gaussian Process, KNN: K-nearest neighbors, DT: Decision tree, BR: Bagging regression, ETR: Extra-trees regression, RFR: Random-forest regression, ABR: AdaBoost Regression, GBR: Gradient boosting regression. Low values of EVS, $\rho, R^2$ indicate that the given features have little influence on the target.}
\label{Table_2}
\begin{tabular}{|c|c|c|c|c|c|c|c|c|c|c|c|}
\hline
       & \multicolumn{11}{c|}{Methods (Mean)}                                                                                            \\ \hline
   & LR         & RR         & SVM     & GP      & KNN     & DT     & BR     & ETR    & RFR    & ABR      & GBR               \\ \hline
$R^2$  & -3.11E+023 & -3.11E+023 & -0.0613 & -0.0205 & -0.188  & -1.02  & -0.111 & -0.129 & -0.11  & -0.00495 & \textbf{0.000347} \\ \hline
$\rho$ & 0.0727     & 0.0727     & 0.104   & 0.024   & 0.00991 & 0.0362 & 0.0769 & 0.0697 & 0.0727 & 0.0649   & \textbf{0.142}    \\ \hline
EVS    & -3.11E+023 & -3.11E+023 & -0.0511 & -0.0176 & -0.185  & -1.01  & -0.107 & -0.126 & -0.107 & 0.00358  & \textbf{0.00338}  \\ \hline
MAE    & 1.14E+10   & 1.14E+10   & 0.742   & 0.743   & 0.802   & 0.981  & 0.777  & 0.782  & 0.775  & 0.759    & \textbf{0.734}    \\ \hline
RMSE   & 2.29E+11   & 2.29E+11   & 0.95    & 0.932   & 1       & 1.31   & 0.972  & 0.98   & 0.971  & 0.924    & \textbf{0.922}    \\ \hline
\end{tabular}

\vspace{0.4cm}

\begin{tabular}{|c|c|c|c|c|c|c|c|c|c|c|c|}
\hline
       & \multicolumn{11}{c|}{Methods (St. Dev.)}                                                                                   \\ \hline
   & LR        & RR        & SVM    & GP     & KNN    & DT     & BR     & ETR    & RFR    & ABR    & GBR             \\ \hline
$R^2$  & 6.32E+023 & 6.31E+023 & 0.0345 & 0.0104 & 0.0402 & 0.14   & 0.0493 & 0.0328 & 0.0432 & 0.015  & \textbf{0.0281} \\ \hline
$\rho$ & 0.0495    & 0.0495    & 0.0431 & 0.0341 & 0.0385 & 0.0546 & 0.0512 & 0.0358 & 0.0522 & 0.0565 & \textbf{0.0487} \\ \hline
EVS    & 6.31E+023 & 6.3E+023  & 0.0333 & 0.01   & 0.04   & 0.141  & 0.0469 & 0.032  & 0.0427 & 0.0105 & \textbf{0.0271} \\ \hline
MAE    & 2.28E+10  & 2.28E+10  & 0.0228 & 0.0229 & 0.0246 & 0.0453 & 0.0248 & 0.0222 & 0.0236 & 0.0207 & \textbf{0.0233} \\ \hline
RMSE   & 4.60E+11  & 4.60E+11  & 0.028  & 0.0235 & 0.029  & 0.0425 & 0.027  & 0.0267 & 0.0288 & 0.0229 & \textbf{0.0266} \\ \hline
\end{tabular}
\end{table*}

\begin{figure*}[h]
    \centering
    \includegraphics[width=0.6\textwidth]{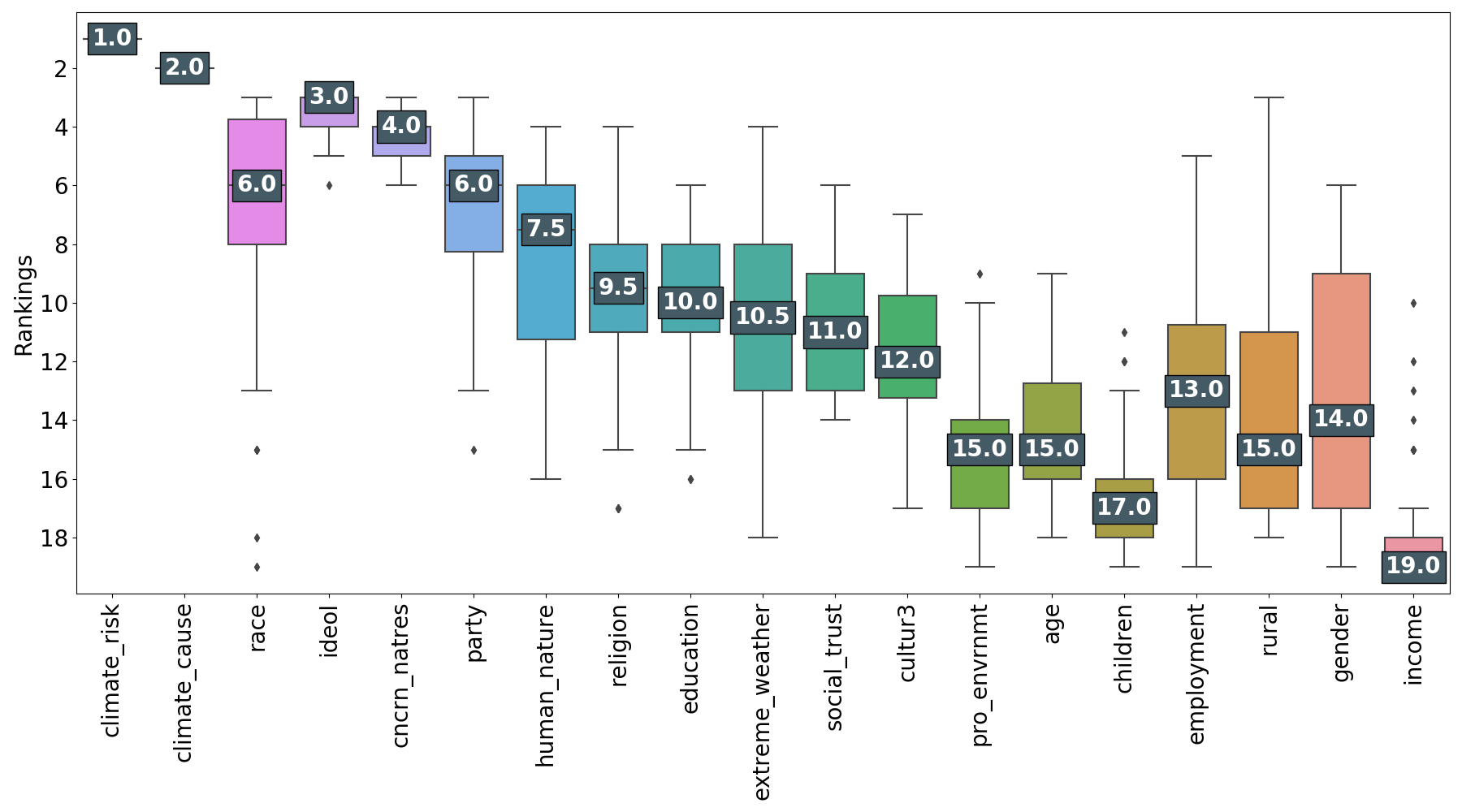}
    \includegraphics[trim={0 0 0 0cm},clip,width=0.35\textwidth]{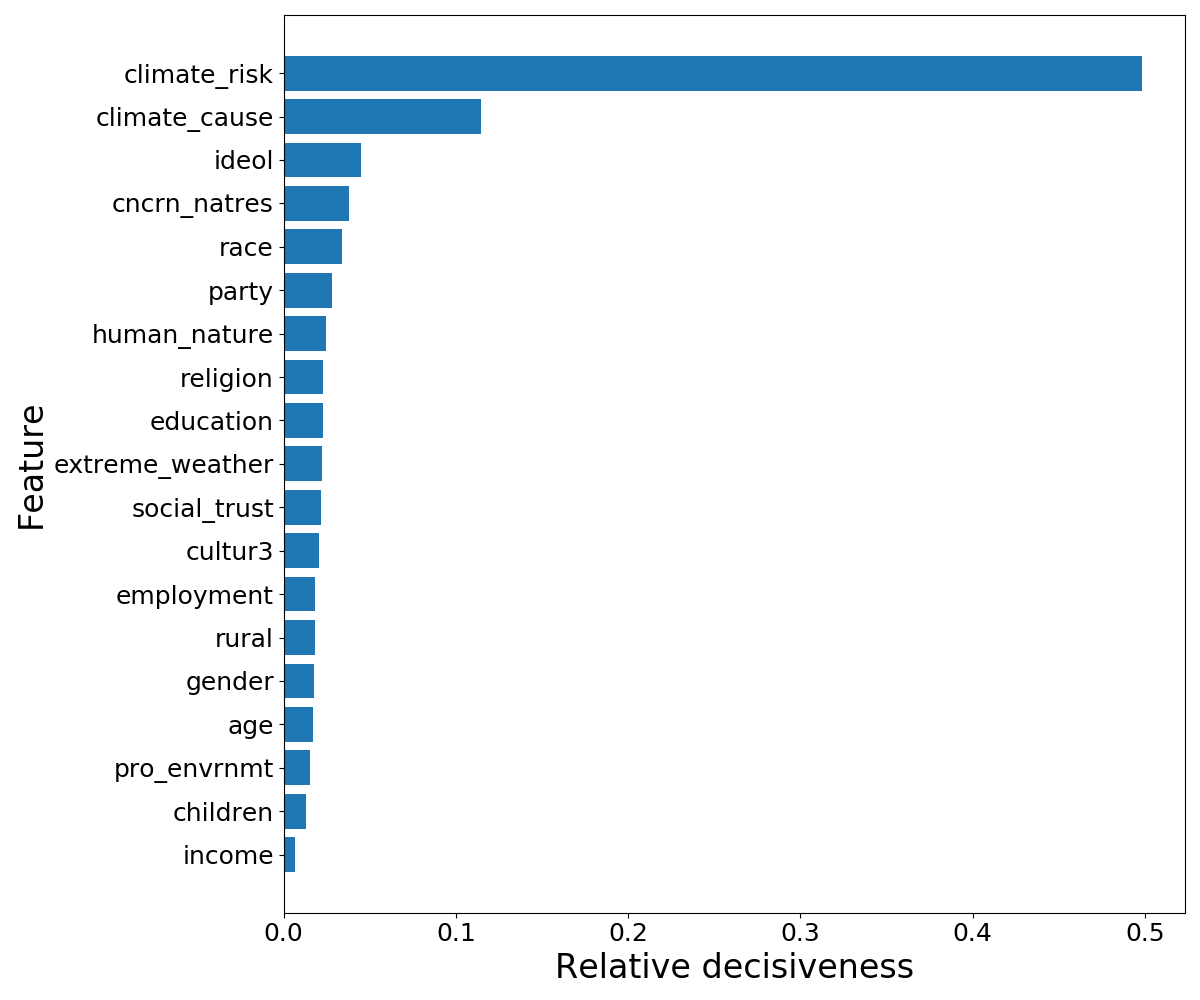}
    \caption{Ensemble feature importance rankings in support of general federal intervention with box-plots (left-showing median ranking in the boxes) and relative importance (right). It is clear, that the variables `climate\_risk' and `climate\_cause' are the most important for explaining this dependent variable. This image corresponds to Table \ref{Table_1}}
    \label{Figure_1}
\end{figure*}

\begin{figure*}[h]
    \centering
    \includegraphics[width=0.6\textwidth]{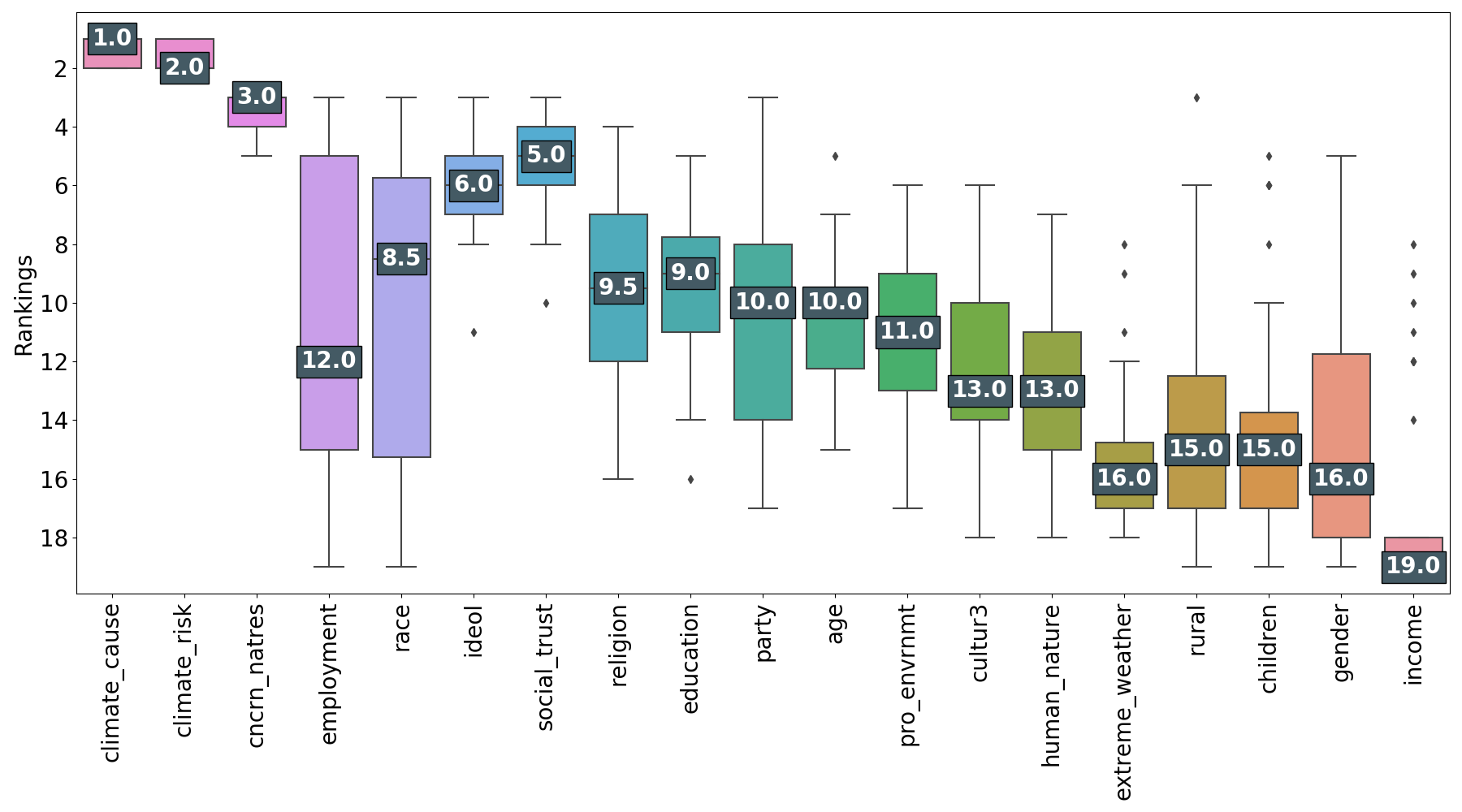}
    \includegraphics[trim={0 0 0 0cm},clip,width=0.35\textwidth]{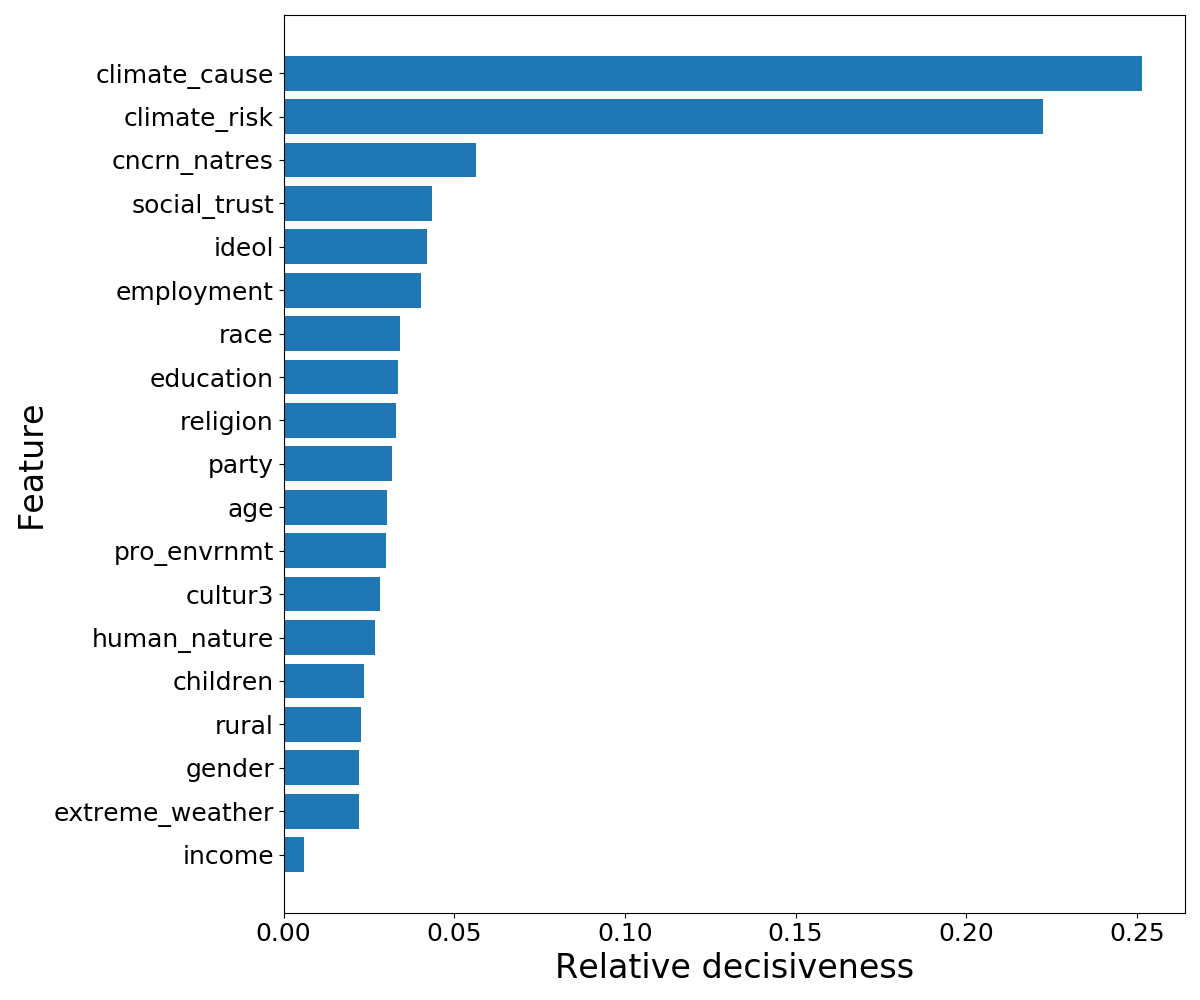}
    \caption{Ensemble feature importance rankings in support of increasing funding for renewable research with box-plots (left-showing median ranking in the boxes) and relative importance (right). It is clear, that the variables `climate\_risk' and `climate\_cause' are the most important for explaining this dependent variable. This image corresponds to Table \ref{Table_3}}
    \label{Figure_2}
\end{figure*}

\begin{figure*}[h]
    \centering
    \includegraphics[width=0.6\textwidth]{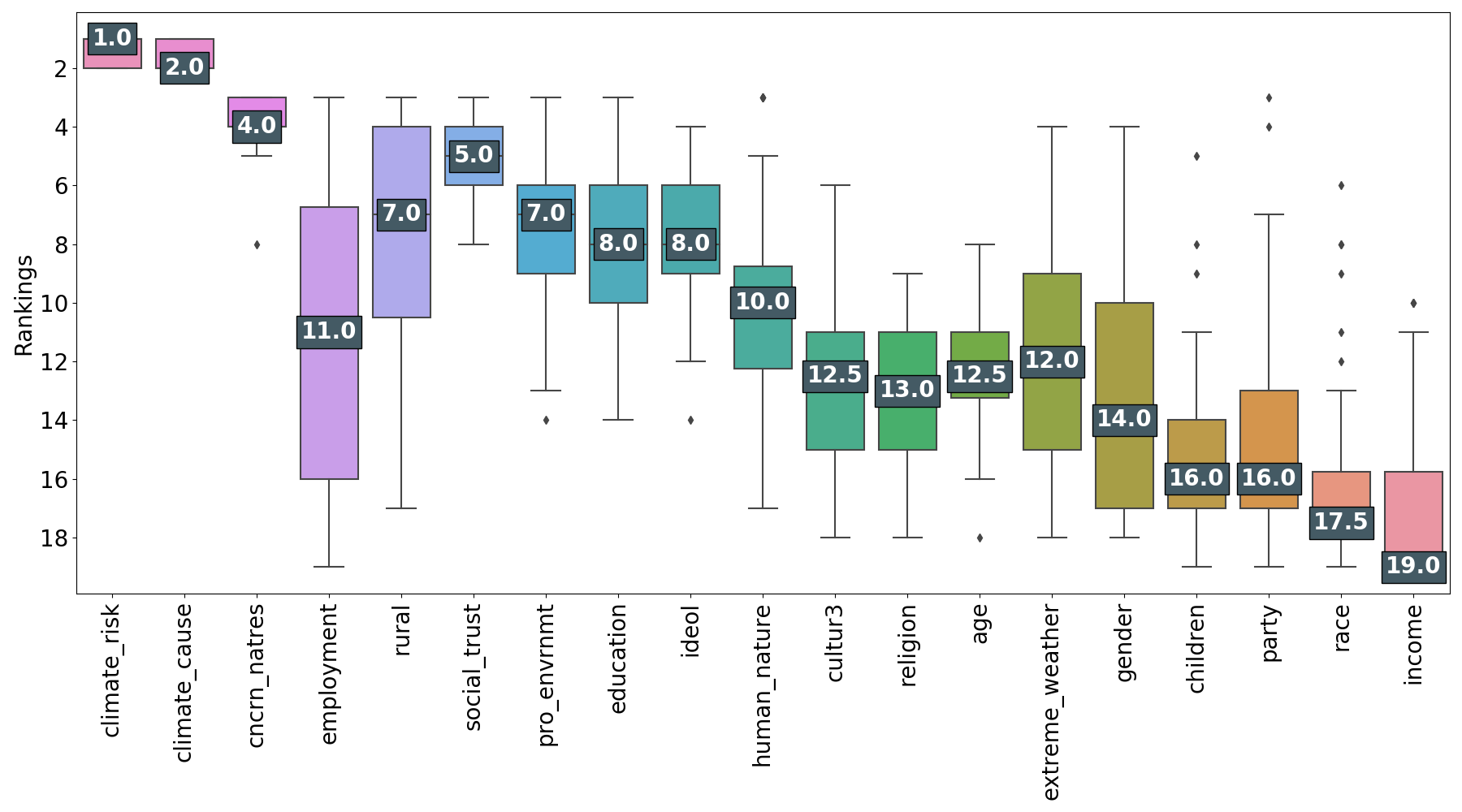}
    \includegraphics[trim={0 0 0 0cm},clip,width=0.35\textwidth]{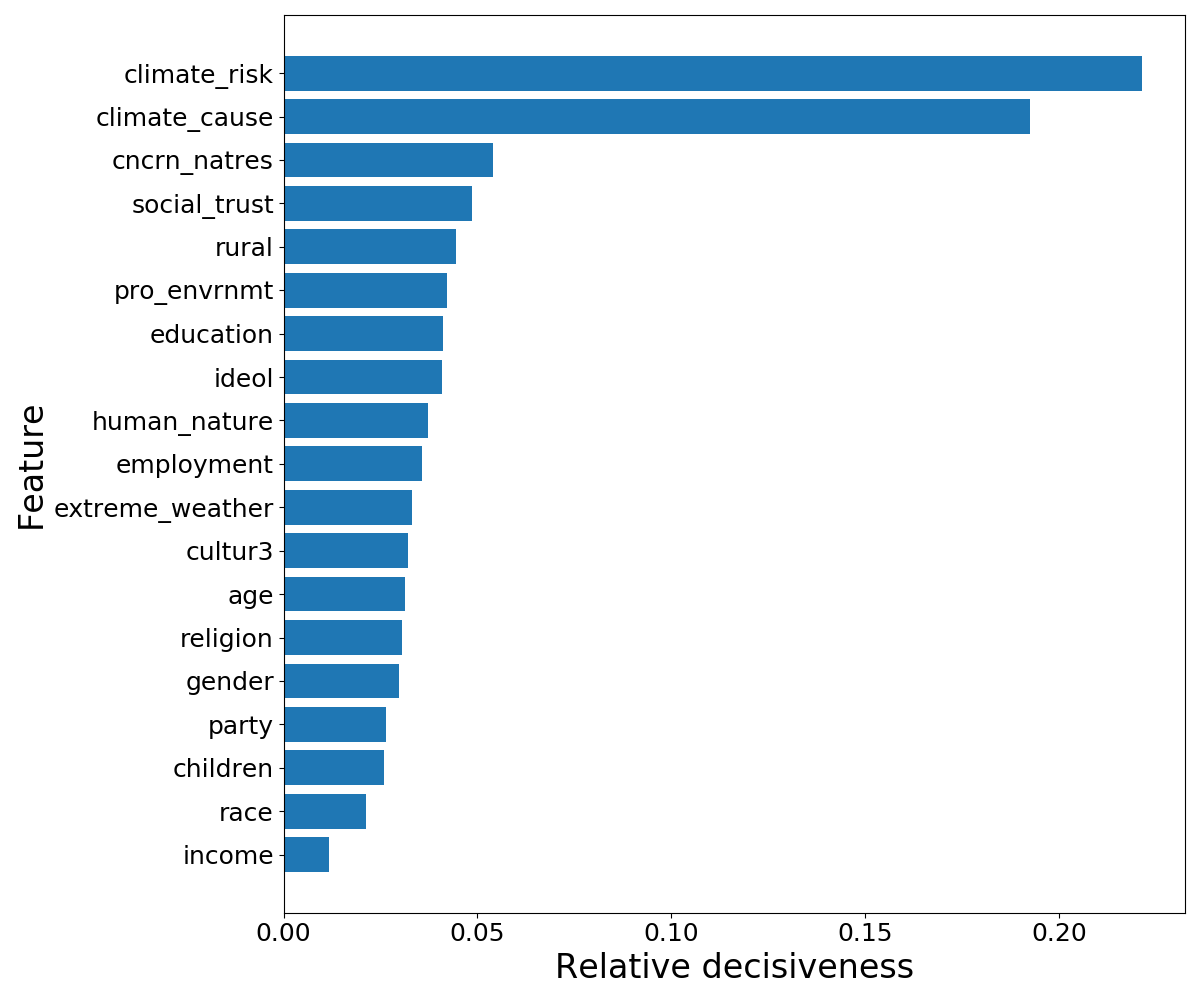}
    \caption{Ensemble feature importance rankings in support of generating renewable energy on public lands with box-plots (left-showing median ranking in the boxes) and relative importance (right). It is clear, that the variables `climate\_risk' and `climate\_cause' are the most important for explaining this dependent variable. This image corresponds to Table \ref{Table_4}}
    \label{Figure_3}
\end{figure*}

\begin{figure*}[h]
    \centering
    \includegraphics[width=0.6\textwidth]{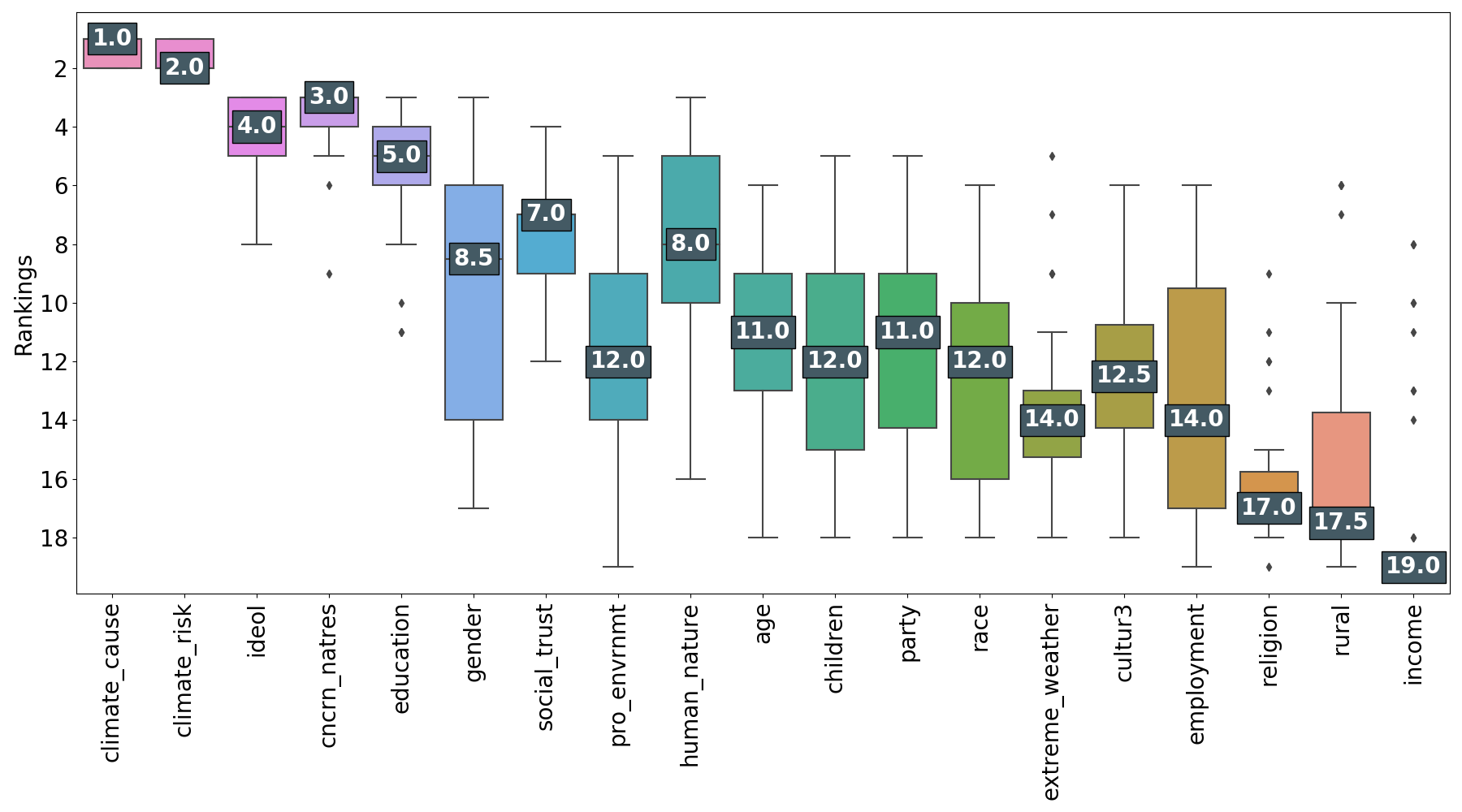}
    \includegraphics[trim={0 0 0 0cm},clip,width=0.35\textwidth]{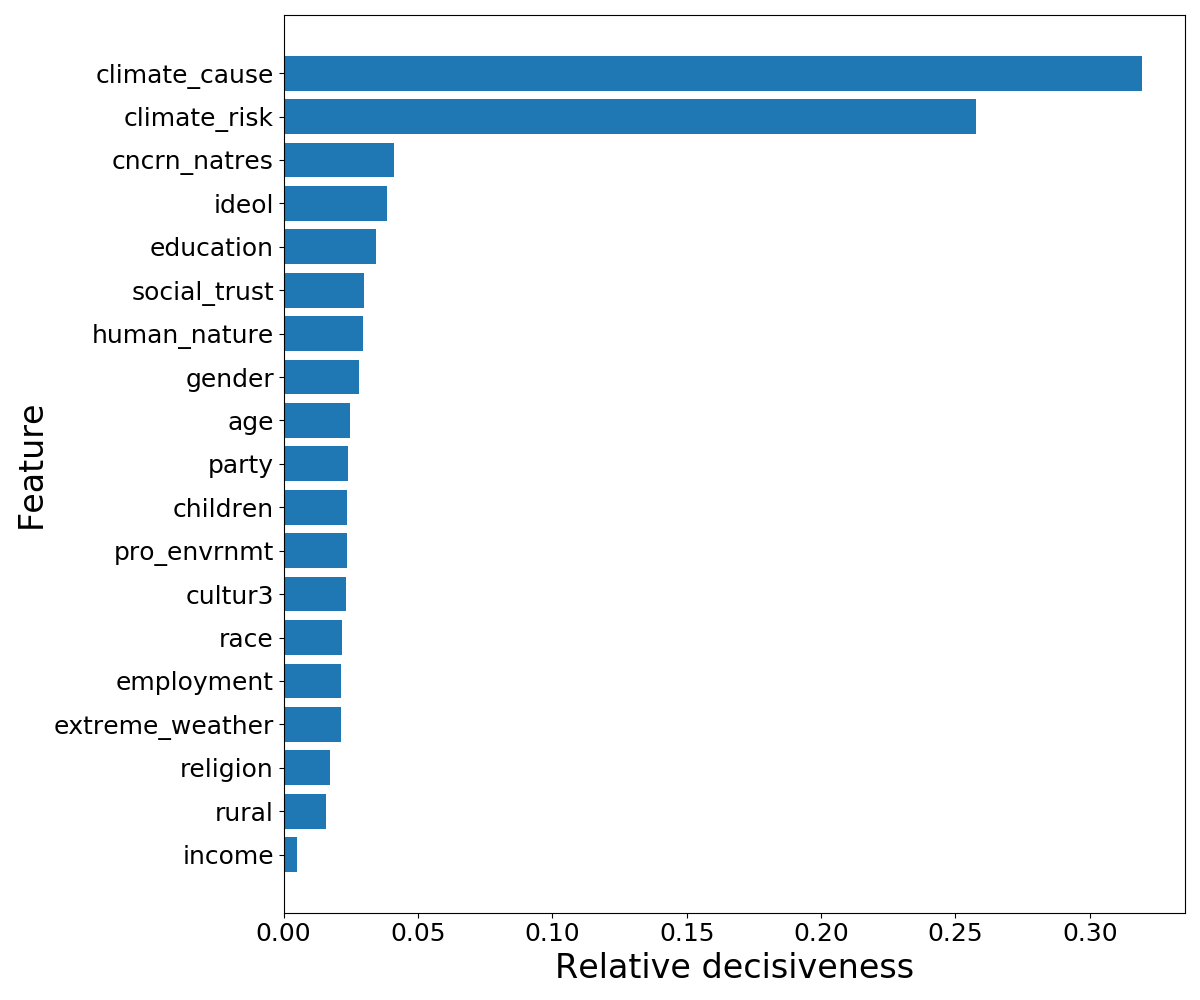}
    \caption{Ensemble feature importance rankings in support of regulating CO2 as a pollutant with box-plots (left-showing median ranking in the boxes) and relative importance (right). It is clear, that the variables `climate\_risk' and `climate\_cause' are the most important for explaining this dependent variable. This image corresponds to Table \ref{Table_5}}
    \label{Figure_4}
\end{figure*}

\begin{figure*}[h]
    \centering
    \includegraphics[width=0.6\textwidth]{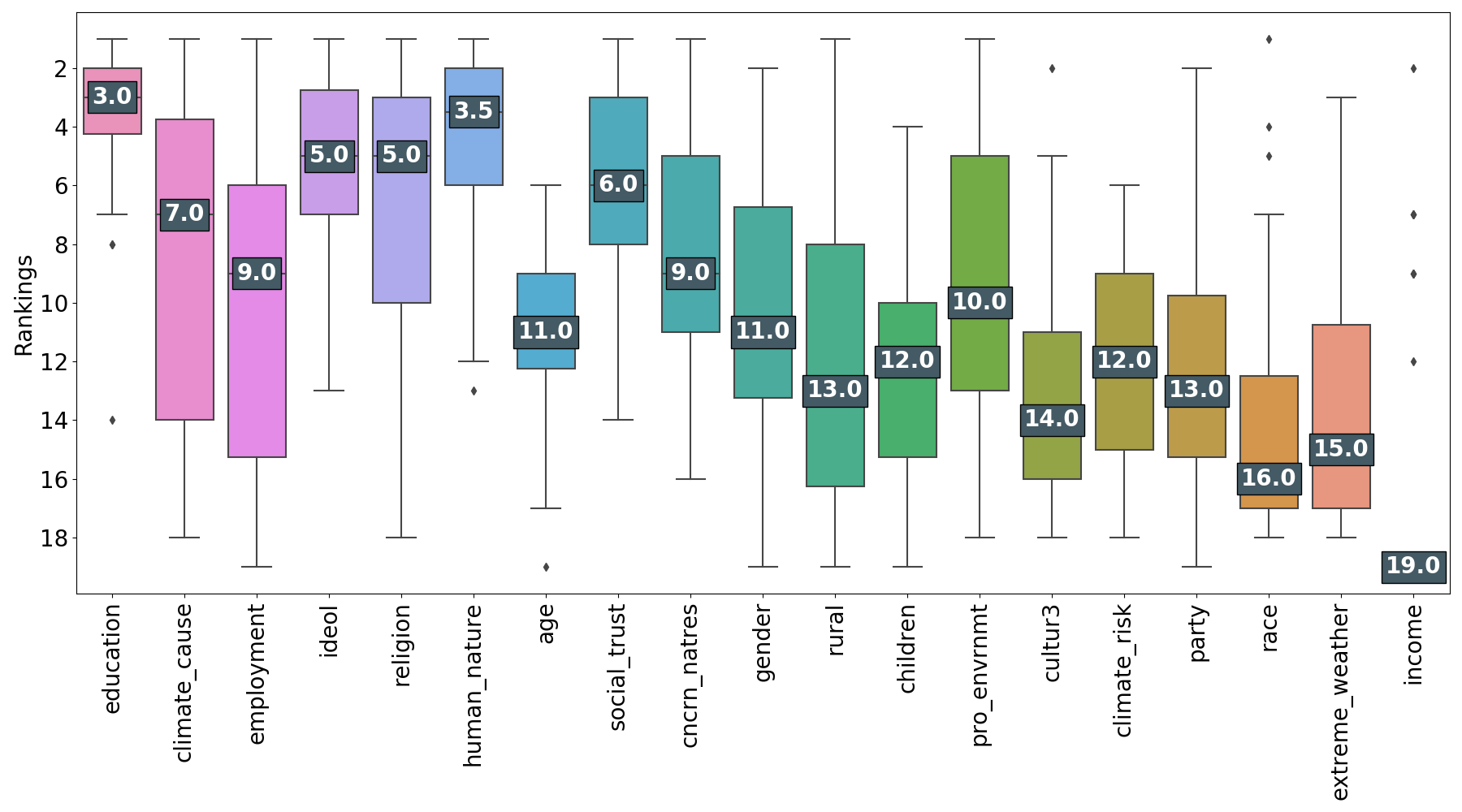}
    \includegraphics[trim={0 0 0 0cm},clip,width=0.35\textwidth]{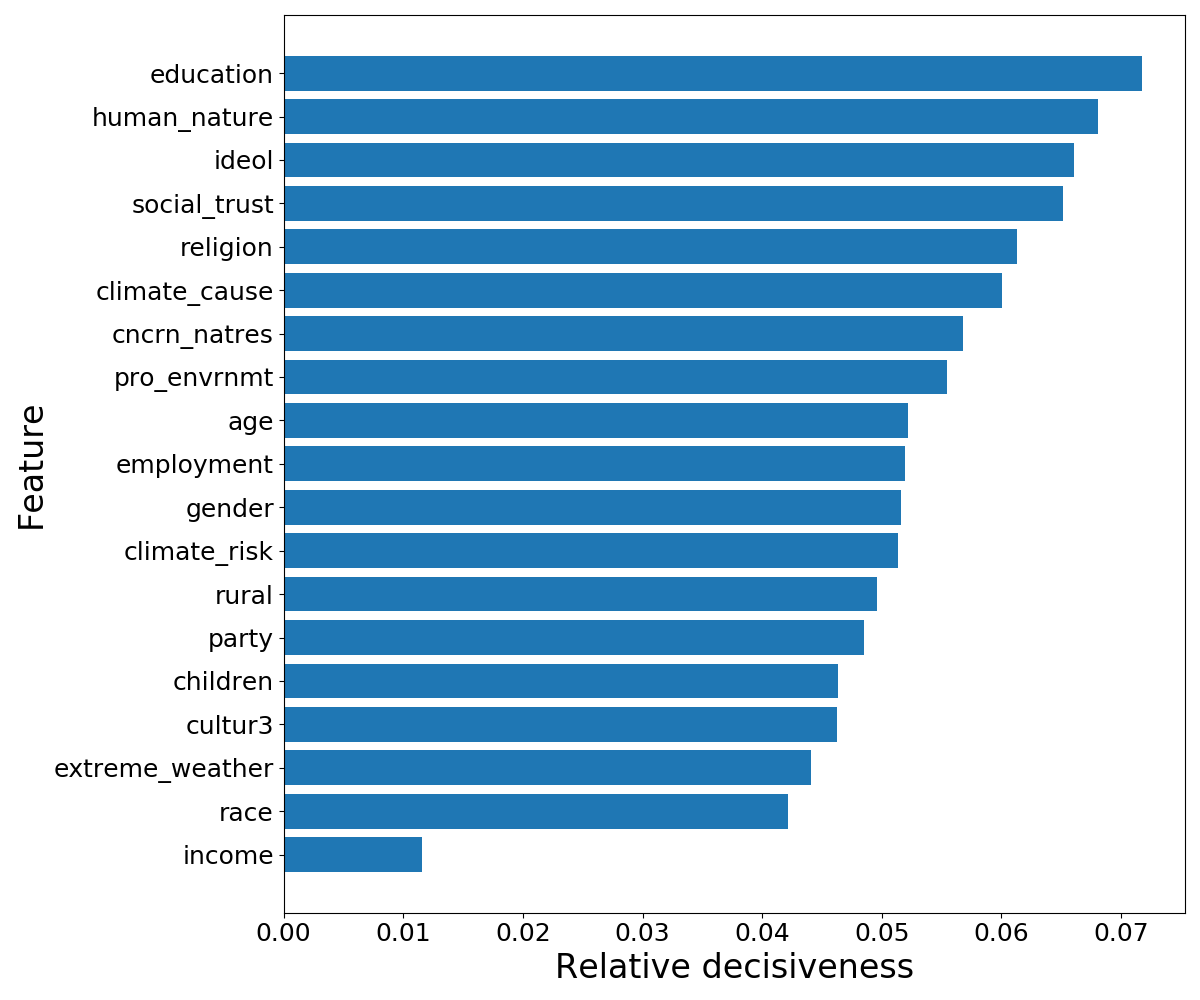}
    \caption{Ensemble feature importance rankings in support of increasing nuclear energy reliance with box-plots (left-showing median ranking in the boxes) and relative importance (right). No clear relationship is observed between features and targets implying special treatment is needed. This image corresponds to Table \ref{Table_2}.}
    \label{Figure_5}
\end{figure*}

After ascertaining that GBR is the best method on the given datasets - we utilize its feature importance extraction capability. We can then determine which of the features are most important for explaining the dependent variable. To that end, we obtain the feature importance values for the 40 different folds (with the 80\%-20\% split of the entire data set as explained previously). Following this procedure, one is left with 40 sets of feature importance. We stack all the features and display box plots which also show median, quartile, minimum, maximum and outlier values as shown in Figures \ref{Figure_1} to \ref{Figure_5}. In addition, variation in importance trends (due to sub-sampling of the data set) is also shown through the use of box plots of rankings. Note that the $x-$axes in all box plots are ordered according to increasing \emph{modal} values of all the feature importance rankings. In other words, the third variable from the left on the $x-$axis had a ranking of 3 the most number of times. Box-plots for each variable in the figures provide more detailed information about the decisiveness of each variable and increase confidence in our conclusions. Also, note that a higher modal value implies \emph{lower} ranking and consequently, a lower importance in our formulation. 

Figure \ref{Figure_1} indicates that the perceived risk associated with climate change is the most decisive factor for this dependent variable. The second most important factor is seen to be related to an individual's belief in greenhouse gas emissions as a contributor to climate change. Other factors are seen to be virtually of the same order of magnitude of importance as shown in the relative importance plots and the amount of variance in their rankings for the different folds. These trends are observed once more in Figure \ref{Figure_2} where the two most important features are once again related to risk perception of climate change and whether greenhouse gases cause climate change. We note, however, that both these variables have a reduced degree of difference in their relative importance (as against the dependent variable of Figure \ref{Figure_1}). Figures \ref{Figure_3} and \ref{Figure_4} show results very similar to Figure \ref{Figure_1} with the same two variables leading the relative importance rankings. This points towards similar levels of influence of climate change risk perception and level of belief in greenhouse gas emissions towards policies influencing funding for renewable energy research, generating renewable energy on public land and regulating CO2 as a pollutant.

The key finding of this article may be observed in Figure \ref{Figure_5} where the dominance of the two aforementioned variables is no longer seen. As seen in the plot related to relative decisiveness, almost all variables are equally important. This is also evidenced by the large variation in rankings for different folds as outlined in the box plots. This suggests that policy initiatives relevant to nuclear energy cannot be clubbed with those related to renewable energy. Essentially, public support for climate change mitigating initiatives may not readily translate to public support for nuclear energy. Note that the low goodness of fit metrics (i.e., $R^2$, $\rho$, etc.) also indicate that this might be an issue of the data set at hand. In either case, there are likely other factors at play, not captured by the data set and survey questions, and that the understanding of support for this policy initiative needs special treatment (see Table \ref{Table_2}).

\section{Conclusions}

Given the importance of public support for policy change and implementation and the urgency of climate change, practitioners and decision-makers must promote individual support for climate change mitigation policy. To analyze the factors that affect individual support the most intimately, in this article, we have utilized nonlinear machine learning methods in place of traditional linear regression techniques. This is because the latter are built on hypotheses, such as uncorrelated features, which are inherently wrong for socio-economic datasets. We have utilized gradient boosted regressors to examine the relative decisiveness of factors previously investigated in structuring public support for climate change mitigation policies. This is performed after a thorough evaluation of multiple data-driven methods using 40-fold cross-validation on the given data set.

To that end, we have looked at public support for general federal government intervention to mitigate climate change risks and four specific climate change mitigation policy options such as increasing funding for renewable research, generating renewable energy on public lands, regulating CO2 as a pollutant and increasing nuclear energy reliance as a climate change mitigation effort. It should be noted that the first three mitigation options are related to renewable energy options, while the fourth, nuclear energy, is considered a non-renewable energy option. However, these four policy options are known to, with varying degrees, mitigate the risks associated with climate change and are often considered equivalent.
 
Our analyses show, first, that the risk perception of climate change and an individual's belief of whether climate change is caused by greenhouse gas emissions are the most decisive factor in shaping public support for general federal government intervention. Furthermore, these features also play the most decisive role in shaping public support for three options related to renewable energy. These results support previous findings in political science, public policy and other social science literature. However, the risk perception of climate change seems to be one of many important factors in shaping individual support for nuclear energy reliance. While nuclear energy is often heavily discussed in the context of climate change solutions, our results indicate that the public may not connect nuclear energy options with climate change issues explicitly. Instead, our analyses indicate that many different variables may play a decisive role, together, to shape individual support for nuclear energy reliance. Additionally, the low accuracy of the predictive models for nuclear energy reliance data, suggests the need for special treatment in future data collection with rephrased questions to explicitly connect climate change mitigation and nuclear energy.

The methods studied in this article have limitations and further study is needed to explore the results shown here. Since the gradient boosting regressor does not provide knowledge on \emph{how} these relatively important variables affect public support for climate change mitigation policy, the direction of explanatory factors is beyond the scope of this paper. For future work, our goal is to introduce a further layer of interpretability using Shapley additive explanations \cite{lundberg2017unified} to ascertain the \emph{direction} of influence of the features on the targets. Given previous research and the findings of this article, it is expected that individuals are more likely to support climate change mitigation policy options that are explicitly related to renewable energy options. However, we expect to find some off-nominal behavior in the way the features influence an individual's support for nuclear energy reliance as shown in this dataset. This article also draws attention to a critical limitation of using conventional data science methods in the social sciences. Machine learning for socio-economic and psychological data sets requires the need for interpretability first and accuracy second. Conventional `goodness-of-fit' such as the $R^2$ find limited use for relationship calculations and may often be misleading.

The results also indicate that there should be some intervention to accentuate public risk perception for general federal government intervention and climate change mitigation policy options related to renewable energy. To do so, scholars might need to understand more about how individuals perceive risks associated with climate change issues. Finally, under circumstances where there have been numerous kitchen-sink models to understand public support for climate change mitigation policy \footnote{Kitchen-sink models refer to statistical regression which includes a long list of possible explanatory variables to explain variations in a dependent variable}, our analyses help us focus our research attention to the variables that are most likely to predict and improve public support for climate change mitigation policy.  Our future work also seeks to apply similar investigations to global data sets.

\section{Acknowledgements}

This material is based upon work supported by the U.S. Department of Energy (DOE), Office of Science, Office of Advanced Scientific Computing Research, under Contract DE-AC02-06CH11357. This research was funded in part and used resources of the Argonne Leadership Computing Facility, which is a DOE Office of Science User Facility supported under Contract DE-AC02-06CH11357. Data were collected by the University of Oklahoma with support from the National Science Foundation under Grant No. IIA-1301789. This paper describes objective technical results and analysis. Any subjective views or opinions that might be expressed in the paper do not necessarily represent the views of the U.S. DOE or the United States Government.

\bibliographystyle{unsrt}
\bibliography{references}  

\end{document}